\documentstyle[iopconf1]{article}
\begin{document}
\title{Electroweak Baryon Number Violation and Constraints on 
Left-handed Majorana Neutrino Masses }

\author{Utpal Sarkar\dag\footnote{E-mail:utpal@prl.ernet.in}}

\affil{\dag Physical Research Laboratory, Ahmedabad - 380 009, INDIA}

\beginabstract

During a large period of time, the anomalous baryon number 
violating interactions are in equilibrium, when the $(B+L)$ 
asymmetry is washed out. If there is any lepton number 
violation during this period, that will also erase the $(B-L)$ 
asymmetry. As a result, survival of the baryon asymmetry of 
the universe pose strong constraints on lepton number violating 
interactions. We review here the constraints on the left-handed 
Majorana neutrino masses arising from this survival requirement
of the baryon asymmetry of the universe. We then briefly review 
models of leptogenesis, where lepton number violation is used to 
generate a  baryon asymmetry of the universe and hence the 
constraints on the Majorana neutrino mass is relaxed. 

\endabstract 

\section{Introduction} 

The generation of the baryon asymmetry of the universe starting from a
symmetric universe is   one   of the  very challenging  question    in
cosmology \cite{kolb}. Sakharov  \cite{sakh} pointed out that for
this purpose  we need three   conditions (A) {\it Baryon
number violation}, (B) {\it $C$ and $CP$ violation}, and (C) {\it 
Departure from thermal equilibrium}. It  was then realised that grand
unified     theories   (GUTs)    satisfies     all    these  criterion
\cite{gutbar,gutrev}.   Because  of   the   quark-lepton  unification
baryon number is  violated in GUTs.  Since fermions belong
to chiral representation, $C$ is maximally violated in GUTs. Violation
of $CP$ was the  only crucial point, which  had to be  incorporated in
these theories. However, it was not difficult  to consider some of the
couplings  to be complex so  that there exist  tree level and one loop
diagrams  which could interfere to  give us enough baryon asymmetry in
the  decays of the  heavy gauge and higgs  bosons. It was noticed that
the   masses    of these   heavy   bosons  are    constrained   by the
out-of-equilibrium condition to be very high.  However, from the gauge
coupling unification also the masses of  these particles were found to
be  around  the  same  value.  In fact,   the proton decay
constrains also require the masses of these particles to be around the
same scale \cite{gutrev}. 

This was considered to be one  of the major successes  of GUTs that it
can explain the baryon asymmetry of the  universe. After several years
it was realised that  the chiral nature  of the weak interaction  also
breaks the global baryon  and lepton numbers in the standard model 
\cite{hooft}. Since these
classical global  baryon  and lepton number   symmetries are broken by
quantum effects due  to the presence  of the  anomaly, these processes
were found   to be very weak  at  the zero temperature. But  at finite
temperature these baryon and lepton number violating interactions were 
found  to be very  strong in the presence of
some static topological field configuration - sphalerons \cite{krs}. 
In fact, these
interactions are so strong that in no time the particles and anti-particles
attain their equilibrium distributions. As a result, since $CPT$ is 
conserved and hence the masses of the particles and anti-particles
are same, the number density of baryons becomes same as that of the
anti-baryons and that wash out any primordial baryon asymmetry of 
the universe. which are generated during the GUT phase transition.
This started renewed interest in the subject of the baryon asymmetry 
of the universe.

Now this problem takes two directions. First, how to generate a 
baryon asymmetry of the universe, and second, which are the interactions
that can wash out the baryon asymmetry of the universe and what constrains
they give us on the various parameters of the particle physics models.
Since the electroweak anomalous processes breaks both the baryon and
the lepton numbers, still conserving the $(B-L)$ quantum number, the
baryon asymmetry of the universe is no longer independent of the 
lepton number violation of the universe \cite{krs,fy1,fy2,ht}. 
If there is very fast lepton
number violation before the electroweak phase transition, then that
can erase the $(B-L)$ asymmetry of the universe and hence the baryon 
asymmetry of the universe. 
On the other hand, if any lepton asymmetry is 
generated at some high temperature, that can get converted to a 
baryon asymmetry of the universe before and during the 
electroweak phase transition.

The first thing then comes to mind to make use of the baryon number 
violation of the standard model to generate a baryon asymmetry of the 
universe. There were several attempts towards this direction 
\cite{ewbar,shapos}. However, in these models one needs to protect
the generated baryon asymmetry after the phase transition, which 
requires the mass of the standard model doublet higgs boson to be
substantially light 
\cite{higgsbound}. The present
experimental limit of 95 GeV on the mass of the higgs
boson {\it almost} rule out all these possibilities. Then the most interesting
scenario remains for the understanding of the baryon number of the
universe is through lepton number violation \cite{fy1,lg1,lg2,lg3}, 
which is also
referred to as leptogenesis. In this 
scenario one generates a lepton asymmetry of the universe, which is
the same as the $(B-L)$ asymmetry of the universe. This $(B-L)$ 
asymmetry of the universe then get converted to the baryon asymmetry of 
the universe during the period when the sphaleron fields maintain
the baryon number violating interactions in equilibrium. On the 
other hand, if there is vary fast lepton number violation in the
universe during this period, that can also wash out the baryon asymmetry
of the universe \cite{fy2,lgb1,lgb2,lgb3}. 

In this article I shall discuss the constraints on the left-handed
Majorana neutrino mass, which comes from the survival of the baryon
asymmetry of the universe. Since, in models of leptogenesis these
constraints are weakened, we shall also discuss models of 
leptogenesis briefly. The limitations of these constraints
will be mentioned. 

\section{Sphaleron processes in thermal equilibrium and 
relation between baryon and lepton numbers}

Anomaly breaks any classical symmetry of the lagrangian at the 
quantum level. So, all local gauge theories should be free of
anomalies. However, there may be anomalies corresponding to any
global current, {\it i.e.}, in the triangle loop while two gauge
bosons would couple to two vertices, a global current will be
associated to the third vertex. 
That will simply mean that such global symmetries
of the classical lagrangian are broken through quantum effects.

In the standard model the chiral nature of the weak interaction
makes the baryon and lepton number anomalous. If
we associate the $SU(2)_L$ or the $U(1)_Y$ gauge bosons at the 
two vertices of a triangle diagram and associate a global current
corresponding to baryon or lepton numbers at the third vertex, 
then sum over all the fermions in the standard model will give
us non vanishing axial current \cite{hooft}
$$ \delta_\mu j^{\mu 5}_{(B+L)} = 6 [ {\alpha_2 \over 8 \pi}
W_a^{\mu \nu} \tilde{W}_{a \mu \nu} +  {\alpha_1 \over 8 \pi}
Y^{\mu \nu} \tilde{Y}_{\mu \nu} ] $$
which will break the $(B+L)$ symmetry. 
However, the anomaly corresponding to
the baryon and lepton number are same and as a result there is
no anomaly corresponding to the $(B-L)$ charge. 

Because of the anomaly \cite{hooft}, baryon and lepton numbers are broken
during the electroweak phase transition, 
$$ \Delta (B+L) = 2 N_g {\alpha_2 \over 8 \pi} \int d^4 x
W_a^{\mu \nu} \tilde{W}_{a \mu \nu} = 2 N_g \nu , $$
but their rate is very
small at zero temperature, since they are suppressed by quantum 
tunnelling probability, $\exp[- {2 \pi \over \alpha_2} \nu] ,$ where
$\nu$ is the Chern-Simmons number.

At finite temperature, however, it has been shown that there exists
non-trivial static topological soliton configuration, 
called the sphalerons, which enhances the baryon number violating 
transition rate \cite{krs}. As a result, at finite
temperature these baryon and lepton number violating processes are no
longer suppressed by quantum tunneling factor, rather the suppression
factor is now replaced by the Boltzmann factor $$ \exp[- {V_0 \over
T } \nu]$$ where the potential or the free energy $V_0$
is related to the mass of the sphaleron field, which is 
about TeV. As a result, at temperatures between 
\begin{equation} 
10^{12} GeV > T > 10^2 GeV  \label{per}
\end{equation}
the sphaleron mediated baryon and lepton number violating
processes are in equilibrium. For the simplest scenario of $\nu = 1$,
the sphaleron induced processes are $\Delta B = \Delta L = 3$, 
given by,
\begin{equation} 
|vac> \longrightarrow [u_L u_L d_L e^-_L + c_L c_L s_L \mu^-_L
+ t_L t_L b_L \tau^-_L] . \label{sph}
\end{equation}
These baryon and lepton number violating fast processes can,
in general, wash out any pre-existing baryon or lepton number asymmetry
of the universe. 
However, if there are any $(B-L)$ asymmetry of the 
universe, that will not be washed out. In fact, any $(B-L)$ asymmetry
before the electroweak phase transition will get converted to a
baryon and lepton asymmetry of the universe during this 
phase transition, which can be seen from an analysis of the
chemical potential \cite{ht}. 

We consider all the particles to be 
ultrarelativistic, which is the case above the electroweak scale.
At lower energies, a careful analysis has to
include the mass corrections, but since they are anyway small
we ignore them for our present discussion. The
particle asymmetry, {\it  i.e.} the difference  between the number of
particles ($n_{+}$) and the number of antiparticles ($n_{-}$) can be
given in terms of the chemical potential of the particle species $\mu$ 
(for antiparticles the chemical potential is $-\mu $) as
\begin{equation}  
n_{+}-n_{-}=n_{d}{\frac{gT^{3}}{6}}\left( {\frac{\mu}{T}}\right), 
\end{equation}  
where $n_{d}=2$ for bosons  and $n_{d}=1$ for fermions. 

In the standard model the quarks and leptons transform under 
$SU(3)_C \times SU(2)_L \times U(1)_Y$ as,
\begin{equation}
\left( 
\begin{array}{c}
u_i \\ 
d_i
\end{array}
\right)_L \sim (3, 2, 1/6), ~~~ u_{iR} \sim (3, 1, 2/3), ~~~
d_{iR} \sim (3, 1, -1/3);
\end{equation}
\begin{equation}
\left( 
\begin{array}{c}
\nu_i \\ 
e_i
\end{array}
\right)_L  \sim (1, 2, -1/2), ~~~e_{iR} \sim (1, 1, -1).
\end{equation}
where, $i = 1,2,3$ corresponds to three generations. 
In addition, the scalar sector consists of the usual Higgs doublet,
\begin{equation}
\left( 
\begin{array}{c}
\phi^+ \\ 
\phi^0
\end{array}
\right) \sim (1, 2, 1/2)
\end{equation}
which breaks the electroweak gauge symmetry $SU(2)_L \times U(1)_Y$
down to $U(1)_{em}$. When these leptons
interact with other particles in equilibrium, the chemical potentials 
get related by simple additive relations, and that will allow us to 
relate the lepton asymmetry $n_L$ to the baryon asymmetry during the
electroweak phase transition. 

During the period $10^{2} GeV < T < 10^{12} GeV$, the sphaleron \cite{krs} 
induced electroweak $B+L$ violating interaction arising due
to the nonperturbative axial-vector anomaly \cite{hooft}
will be in equilibrium along with the other interactions.  In  Table 1, 
we present all other interactions and the corresponding relations between  
the chemical potentials.  In the third column we give the chemical potential 
which we eliminate using the given relation.  We start with chemical 
potentials of all the quarks ($\mu _{uL},\mu _{dL},\mu _{uR},\mu _{dR}$); 
leptons ($\mu _{aL},\mu _{\nu aL},\mu _{aR}$,  where 
$ a=e,\mu,\tau $); gauge bosons ($\mu _{W}$  for $W^{-}$, 
and 0 for all others); and the Higgs scalars ($\mu _{-}^{\phi  },
\mu _{0}^{\phi}$).  

\begin{table}[htb]
\caption {Relations among the chemical potentials}
\begin{center}
\begin{tabular}{||c|c|c||}
\hline \hline
Interactions& $\mu$ relations&$\mu $ eliminated \\
\hline
{$D_{\mu }\phi ^{\dagger}D_{\mu
}\phi $}&{$\mu _{W}=\mu _{-}^{\phi }+\mu _{0}^{\phi
}$}&{$\mu_{-}^{\phi }$}\\
{$\overline{q_{L}}\gamma _{\mu}q_{L}W^{\mu }$}&{$\mu
_{dL}=\mu _{uL}+\mu _{W}$}&{$\mu_{dL}^{{}}$}\\
{$\overline{l_{L}}\gamma _{\mu }l_{L}W^{\mu
}$}&{$\mu_{iL}^{{}}=\mu _{\nu iL}^{{}}+\mu
_{W}$}&{$\mu_{iL}$}\\
{$\overline{q_{L}}u_{R}\phi ^{\dagger
}$}&{$\mu_{uR}=\mu _{0}+\mu_{uL}$}&{$\mu_{uR}^{{}}$}\\
{$\overline{q_{L}}d_{R}\phi $}&{$\mu
_{dR}=-\mu_{0}+\mu _{dL}$}&{$\mu_{dR}$}\\
{$\overline{l_{iL}}e_{iR}\phi
$}&{$\mu _{iR}^{{}}=-\mu_{0}+\mu _{iL}^{{}}$}&{$\mu_{iR}$}\\
\hline \hline
\end{tabular}
\end{center}
\end{table}

The chemical potentials of the neutrinos always enter as a sum 
and for that reason we can consider it as one parameter.
We can then express all the chemical potentials in terms of the following
independent chemical potentials only,
\begin{equation}
\mu _{0}=\mu _{0}^{\phi };~~\mu _{W};~~\mu _{u}=\mu _{uL};~~
\mu = \sum_i \mu _{i}= \sum_i \mu _{\nu iL}.
\end{equation}
We can further eliminate one of these four potentials by making use of the
relation given by the sphaleron processes (\ref{sph}).  
Since the sphaleron interactions
are in equilibrium, we can write down the following $B+L$ violating relation
among the chemical potentials for three generations,
\begin{equation}
3\mu _{u}+2\mu _{W}+ \mu =0.
\end{equation}
We then express the baryon number, lepton numbers and the
electric charge and the hypercharge number densities in terms of these
independent chemical potentials,
\begin{eqnarray}
B &=&12\mu _{u}+6\mu _{W} \\
L_{i} &=&3\mu +2\mu _{W}-\mu _{0}  \\
Q &=&24 \mu _{u}+(12+2m)\mu _{0}-(4+2m)\mu _{W} \\
Q_{3} &=&-(10+m)\mu _{W} 
\end{eqnarray}
where $m$ is the number of Higgs doublets $\phi$.

At temperatures above the electroweak phase transition, $T>T_{c}$, both $Q$
and $Q_{3}$ must vanish. These conditions and the
sphaleron induced $B-L$ conserving, $B+L$ violating condition can be
expressed as
\begin{eqnarray}
<Q>=0 &\Longrightarrow &\mu _{0}=\frac{-12}{6+m}\mu_u \\   
<Q_{3}>=0 &\Longrightarrow &\mu _{W}=0 \\
{\rm Sphaleron~~ transition } &\Longrightarrow &\mu _{a}=- 5 \mu _{u}
\end{eqnarray}
Using these relations we can now write down the baryon number, lepton
number, and their combinations in terms of the $B-L$ number density as,
\begin{eqnarray}
B &=&\frac{24+4m}{66+13m}~(B-L) \\
L &=&\frac{-42-9m}{66+13m}~(B-L) \\
B+L &=&\frac{-18-5m}{66+13m}~(B-L)
\end{eqnarray}

Below the critical temperature, $Q$ should vanish since the universe is
neutral with respect to all conserved charges. However, since $SU(2)_L$
is now broken we can consider $\mu_0^\phi =0$ and $Q_3 \neq 0$. This 
gives us,
\begin{eqnarray}
<Q>=0 &\Longrightarrow &\mu _{W}=\frac{12}{2+m}\mu_u \\   
\langle \phi \rangle \neq 0 &\Longrightarrow &\mu _{0}= 0 \\
{\rm Sphaleron~~ transition } &\Longrightarrow &\mu _{a}=-3 \mu_W- {
9 \over 2} \mu _{u}
\end{eqnarray}
which then let us write the baryon and lepton numbers as some
combinations of $B-L$ as
\begin{eqnarray}
B &=&\frac{32+4m}{98+13m}~(B-L) \\
L &=&\frac{-66-9m}{98+13m}~(B-L) \\
B+L &=&\frac{-34-5m}{98+13m}~(B-L)
\end{eqnarray}
Thus the baryon and lepton number asymmetry of the universe after 
the electroweak phase transition will depend on the primordial $(B-L)$ 
asymmetry of the universe. If there are very fast lepton number 
violation during the period (\ref{per}), 
that would erase the $(B-L)$ asymmetry of the universe,
and hence we will be left with a baryon symmetric universe at the 
end \cite{fy2,ht}. 
On the other hand, if there is enough lepton asymmetry of the
universe at some high temperature, then that will get converted to a
baryon asymmetry of the universe \cite{fy1,ht}. 

Before proceeding further, we shall briefly discuss what do we mean
when we say that some interaction is fast and that will erase some
asymmetry \cite{kolb,sakh,fry}. In equilibrium the number density of 
particles with non-zero charge $Q$ 
would be same as the antiparticle number density since 
the expectation value of the conserved charge vanishes.
A mathematical formulation of this statement 
reads that the expectation value of any conserved charge $Q$ is 
given by,
$$ <Q> = \frac{{\rm Tr} \left[ Q e^{-\beta H}\right]}{{\rm Tr} \left[ 
e^{-\beta H}\right]} 
$$
and since any conserved charge $Q$ is odd while $H$ is even 
under ${ CPT}$ transformation this expectation value vanishes. 
So for the generation of the baryon asymmetry of the universe 
we have to circumvent this theorem either by including nonzero 
chemical potential, or go away from equilibrium or violate 
${ CPT}$. In most of the popular models ${ CPT}$ 
conservation is assumed and one starts with vanishing chemical 
potential for all the fields which ensures that the entropy is 
maximum in chemical equilibrium. Then to generate the baryon 
asymmetry of the universe one needs to satisfy the 
out-of-equilibrium condition \cite{kolb,sakh,fry}.

The requirement for the out-of-equilibrium condition may also 
be stated in a different way \cite{kolb}. If we assume that the 
chemical potential associated with $B$ is zero and ${ CPT}$ is 
conserved, then in thermal equilibrium the phase space density 
of baryons and antibaryons, given by $[1 + exp(\sqrt{p^2 + 
m^2}/kT)]^{-1}$ are identical and hence there cannot be any baryon 
asymmetry. 

Whether a system is is equilibrium or not can be understood by solving
the Boltzmann equations. But a crude way to put the out-of-equilibrium
condition is to say that the universe expands faster than some 
interaction rate. For example, if some B-violating interaction is
slower than the expansion rate of the universe, this interaction may
not bring the distribution of baryons and antibaryons of 
the universe in equilibrium. In other words, before the 
chemical potentials of the two states gets equal, they move apart from
each other. Thus we may state the out-of-equilibrium condition as
\begin{equation}
\Gamma < \sqrt{1.7 g_*} {T^2 \over M_P}
\end{equation}
where, $\Gamma$ is the interaction rate under discussion, $g_*$ is the 
effective number of degrees of freedom available at that temperature $T$, 
and $M_P$ is the Planck scale.

\section{Bounds on Left-handed neutrino mass}

In this section we shall discuss the constraints on the left-handed 
neutrino mass arising from the constraints of baryogenesis. If there
is Majorana type interactions, which explains the masses of the neutrinos,
then that gives a measure of lepton number violation. If this lepton
number violation is too large before the electroweak phase transition
is over, then that can erase all L asymmetry and hence B asymmetry 
of the universe. 

The first attempt to constraint the neutrino mass was made in a
fairly general framework \cite{fy2}. In the standard model there is 
no lepton number violation. However, one can consider a higher 
dimensional effective operator which violates $(B-L)$, given by
\begin{equation}
L = {2 \over M} l_L l_L \phi \phi + h.c.
\end{equation}
There is no origin of such interactions within the standard model.
So one expects that some new interaction at some high energy 
will give us this effective
interaction at low energy. The scale of the new interaction $M$, which
gives us this interaction, is also the scale of lepton
number violation in this scenario. 

If this interaction is strong enough, it can bring the neutrinos 
in thermal equilibrium with the physical higgs scalars, which can
wash out any lepton asymmetry of the universe. The survival of the
baryon asymmetry of the universe will then require this interaction 
to be slower than the expansion rate of the universe,
\begin{equation}
\Gamma_{L \neq 0} \sim {0.122 \over \pi}{T^3 \over M^2} < 
1.7 \sqrt{g_*} {T^2 \over M_P} \hskip .5in {\rm at}~ T \sim 100 GeV
\end{equation}
which gives a bound on the mass of the heavy scale to be,
$M > 10^9 GeV$. When the higgs doublets $\phi$ acquires a $vev$, the
higher dimensional operator will induce a Majorana mass of the
left-handed neutrinos. This 
bound on the heavy scale $M$ will then imply,
$$ m_\nu < 50 keV  . $$ 

In specific models one may give stronger bounds on the mass of the 
neutrinos \cite{lgb1}. In models with right
handed neutrinos ($N_{Ri},i=e,\mu,\tau$), 
the neutrino masses comes from the see-saw
mechanism \cite{seesaw}. The lagrangian for the lepton sector containing the 
mass terms of the singlet right handed neutrinos $N_i$ and the
Yukawa couplings of these fields with the light leptons is
given by,

\vbox{
\begin{eqnarray}
{\cal L}_{int} & = & \sum_{i} M_i [\overline{(N_{Ri})^c} N_{Ri} + 
   \overline{N_{Ri}} (N_{Ri})^c] \nonumber \\
 & +&  \sum_{\alpha, i} \, h^\ast_{\alpha i} \, \overline{N_{Ri}} 
 \, \phi^{\dagger} \,
     \ell_{L \, \alpha} + \sum_{\alpha, i} h_{\alpha i} \, 
      \overline{\ell_{L \, \alpha}} \phi \, N_{Ri} \\ 
 & +&  \sum_{\alpha, i} \, h^\ast_{\alpha i} \, \overline{(\ell_{L 
 \, \alpha}})^c \, \phi^\dagger \, (N_{Ri})^c + \sum_{\alpha, i} 
 h_{\alpha i} \overline{(N_{Ri})^c} \, 
       \phi \, (\ell_{L \, \alpha})^c  \nonumber
\end{eqnarray}}

\noindent where  $\phi^T =  (-\overline{\phi^\circ},  \phi^-)
\equiv (1,2,-1/2)$ is the usual higgs
doublet of the  standard  mode;  $l_{L \alpha}$  are the
light leptons,  $h_{\alpha i}$ are the complex  Yukawa  couplings
and $\alpha$ is the generation  index.  
Without  loss  of  generality  we work in a basis  in  which  the
Majorana mass matrix of the right handed neutrinos is real and 
diagonal with eigenvalues $M_i$.

Once the higgs doublet $\phi$ acquires a $vev$, the masses of
the neutrinos in the basis $[\nu_{L \alpha} ~~~ N_{Ri}]$ is given by,
\begin{equation}
{\cal M}_\nu = \pmatrix{0 & m \cr m & M}
\end{equation}
where, $m \equiv h_{\alpha i} <\phi> $ and $M \equiv M_{ij}$ 
are $3 \times 3$ matrices. In the limit when all eigenvalues of $M$ are
much heavier than those of $m$, and the matrix $M$ is not singular, 
this matrix may be block diagonalised. It then gives three heavy
right handed Majorana neutrinos with masses $\sim M$ and the 
Majorana mass matrix of the left-handed neutrinos will be given
by,
\begin{equation}
m_\nu = m {1 \over M} m^T .
\end{equation}
In this scenario the see-saw masses of the left-handed neutrinos 
explain naturally why
they are so light. In addition, the decay of the heavy right handed
Majorana neutrinos can generate enough lepton asymmetry of the universe,
which can then get converted to a baryon asymmetry of the universe,
which we shall discuss in the next section.

The decay of $N_{Ri}$ into a lepton and an antilepton,
\begin{eqnarray}
  N_{Ri}  &\to& \ell_{jL} + \bar{\phi}, \nonumber \\
   &\to&  {\ell_{jL}}^c + {\phi} .\label{N}
\end{eqnarray}
breaks lepton number. Since the lightest of the right handed neutrinos
(say $N_1$) will decay at the end, this interaction 
($N_1$ decay) should be slow enough so as
not to erase the baryon asymmetry of the universe, which now implies
\begin{equation}
{|h_{\alpha 1}|^2 \over 16 \pi} M_1 < 1.7 \sqrt{g_*} {T^2 \over M_P}
\hskip .5in {\rm at}~~T = M_1
\end{equation}
which can then give a very strong bound \cite{lgb1} on the mass of the
lightgest of the left-handed neutrinos to be 
$$ m_\nu < 4 \times 10^{-3} eV . $$

In models \cite{oth1,oth2,triplet}, 
where the left-handed  neutrino mass is
not related to any heavy  neutrinos  through  see-saw  mechanism, 
the abovementioned bounds may not be valid.
In addition, there are several  specific
cases even  within the  framework  of see-saw  models,  where these
bounds are not applicable. These bounds are also not valid if some global
U(1) symmetry is exactly conserved up to an electroweak  anomaly
\cite{lgb2}. Furthermore, in some very specific models 
where a baryon asymmetry of the
universe is generated after the electroweak phase transition \cite{mm},
or there are some extra baryon number carrying singlets which decays 
after the electroweak phase transition \cite{sacha}, 
it is possible to avoid all the 
bounds from constraints of survival of the baryon asymmetry of 
the universe. This issue will be discussed in another talk in 
this meeting \cite{sacha} in details.

\begin{figure}[htb]
\mbox{}
\vskip 4.5in\relax\noindent\hskip -.6in\relax
\includegraphics{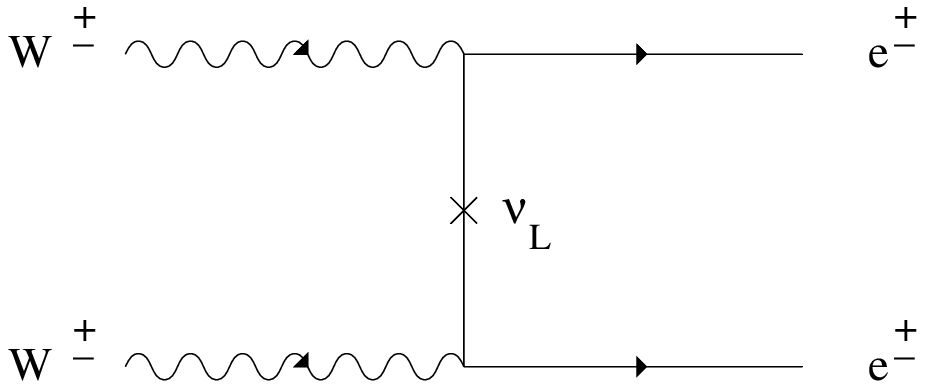}
\vskip -2in
\caption{ Lepton number violating processes $W^{\pm} +
W^{\pm} \to e^{\pm} + e^{\pm}$ mediated by the left handed Majorana
neutrinos.}
\end{figure}

Since some of the bounds derived indirectly from an bound on the scale
of lepton number violation may be circumvented in some models,
we now try to
discuss the question if one can  constrain  the Majorana mass of
the  left-handed  neutrinos  directly.  This  is  possible  since
during the  electroweak  phase  transition,  after the higgs
doublets  acquires a vacuum  expectation  value  ($vev$)  and the
$SU(2)_L$ group is broken, there is no symmetry which can prevent
the mass of the  left-handed  neutrinos.  So if lepton  number is
broken before the  electroweak  symmetry,  then as soon as $\phi$
acquires  a  $vev$  the  left-handed  neutrinos  will  get a mass
$(m_\nu)$.  This  can,  in  principle, induce very fast lepton number
violating  processes, which can wash out any primordial
$(B-L)$  asymmetry and hence the baryon asymmetry of the universe.

There are several lepton
number violating processes, which are active at the time of 
electroweak phase transition. The process,
\begin{equation}
W^+ + W^+ \rightarrow e^+_i + e^+_j \hskip .3in {\rm and}
\hskip .3in W^- + W^- \rightarrow e^-_i + e^-_j
\end{equation}
mediated by a virtual left-handed neutrino exchange as shown
in figure 1 is a lepton number violating interaction active
during the electroweak phase transition. 
Here $i$ and $j$ are the generation indices. Depending
on the physical mass (and also on the elements of the mass matrix)
of the left-handed Majorana neutrinos these processes can wash
out any baryon asymmetry between the time when the higgs
acquires a $vev$ and the $W^\pm$ freeze out, {\it i.e.}, between
the energy scales 250 GeV and 80 GeV. The condition that these 
processes will be slower than the expansion rate of the universe,
\begin{equation}
\Gamma (WW \to e_i e_j) = \frac{\alpha_W^2 
{(m_\nu)}^2_{ij} T^3}{ m_W^4} < 1.7 \sqrt{g_*}
\frac{T^2}{M_p} \hskip .5in {\rm at} \:\:\: T = M_W
\end{equation}
gives a bound on the Majorana mass of the left-handed neutrinos
to be,
\begin{equation}
{(m_\nu)}_{ij} < 20 keV .
\end{equation}
This bound is on each and every element 
of the mass matrix and not on the physical states and 
independent on the existence of any right handed neutrinos.

In general, a Majorana particle can be described by a four component real
field,
\begin{equation}
\Psi_M = \sqrt{m_\nu \over E_\nu} [u_\nu (b_\nu + d_\nu^*) 
{\rm e}^{(-ip.x)} + v_\nu (b_\nu^* + d_\nu) {\rm e}^{(ip.x)} ]
\end{equation}
and hence the charged  current  containing  a Majorana  field, $$
j_\mu =  \overline{\Psi}_l  \gamma_\mu  (1 - \gamma_5)  \Psi_M $$
will have a lepton number  violating  part.  However, this lepton
number  violating  contribution  will always be  suppressed  by a
factor  $(m_\nu/E_\nu)$ and hence the rate of such processes will
be suppressed by a factor $(m_\nu/E_\nu)^2$.  Thus even the decay
of the $W^\pm$ to $e$ and $\nu$ will have lepton number violation
at a rate,
$$ \Gamma (W \rightarrow e \nu) = \frac{\alpha_W}{4} \frac{m_\nu^2 
M_W^2}{T^2 (T^2 + M_W^2)^{1/2}} . $$
The survival of baryon asymmetry of the universe after the electroweak
phase transition again requires this process to be slow enough,
$$ \Gamma(W \to e \nu) < H . $$ This translates to a bound on the Majorana 
mass of the left handed neutrino,
\begin{equation}
m_\nu < 30 \hskip .2in {\rm keV} .
\end{equation}

Similarly,  the decay of the higgs  doublet to an electron and an
anti-neutrino    will   also   have   lepton   number   violating
contribution,  but they will be suppressed by the Yukawa coupling
constants and cannot give stronger bounds.  Similarly  scattering
processes  involving  the higgs, like $\phi + \phi \to l_i + l_j$
(mediated by a virtual  left-handed  neutrino) will contribute to
the evolution of the lepton number  asymmtry of the universe, but
it will  be  much  suppressed  compared  to the  charged  current
interactions and hence cannot give stronger bound to the Majorana
mass of the left-handed neutrinos.

\section{Bounds on neutrino mass in models of Leptogenesis}

In this section we shall discuss two scenarios of leptogenesis.
In the first, one starts
with the standard model and add to it three right handed neutrinos
\cite{fy1,lg1,lg2,lg3}. In total this will add 12 degrees of 
freedom. In the other scenario \cite{triplet}
one adds two complex $SU(2)_L$ triplet 
higgs scalars ($\xi_a \equiv (1,3,-1); a = 1,2$), which also adds 
in total 12 degrees of freedom. At present these two scenarios are
indistinguishable, except that the particle contents are different.
In the left-right symmetric models \cite{lr1,lr2} both these scenarios 
are present, which we shall not discuss here. 
The $CP$ violation required to generate a
lepton asymmetry of the universe are different for the two
scenarios. 

In the standard model neutrinos are massless. In models with right
handed neutrinos the left-handed neutrinos acquire a Majorana mass
through see-saw mechanism as we mentioned earlier. In the triplet
higgs scenario \cite{triplet,trip}
the $vev$s of the triplet higgses can give small
Majorana masses to the neutrinos through the interaction
\begin{equation}
f_{ij} [\xi^0 \nu_i \nu_j + \xi^+ (\nu_i l_j + l_i \nu_j)/\sqrt 2 
+ \xi^{++} l_i l_j] + h.c.
\end{equation}
If the triplet higgs acquires a 
$vev$ and break lepton number spontaneously \cite{gr}, 
then there will be 
Majorons in the problem. Then from the precision measurements of the
$Z-$width that model is ruled out \cite{lep}. However, in a variant of this
model \cite{triplet}
lepton number is broken explicitly through an interaction
of the triplet with the higgs doublet
\begin{eqnarray}
V &=&  \mu (\bar \xi^0 \phi^0 \phi^0 + \sqrt 2 \xi^- \phi^+ \phi^0 + \xi^{--} 
\phi^+ \phi^+) + h.c.
\end{eqnarray}
Let $\langle \phi^0 \rangle = v$ and $\langle \xi^0 \rangle = u$, then the 
conditions for the minimum of the potential relates the $vev$ of the 
two scalars by
\begin{equation}
u \simeq {{-\mu v^2} \over M^2}, \label{min}
\end{equation}
where $M$ is the mass of the triplet higgs scalar.
This is analogous to the usual seesaw mechanism for obtaining small 
Majorana neutrino masses, except that here we do not have any right-handed 
neutrinos. 

Another way of handling the heavy Higgs triplet is to integrate it out.  From 
the couplings of the triplet scalar, 
we obtain the effective nonrenormalizable term
\begin{equation}
{{-f_{ij} \mu} \over {M^2}} [\phi^0 \phi^0 \nu_i \nu_j - \phi^+ \phi^0 
(\nu_i l_j + l_i \nu_j) + \phi^+ \phi^+ l_i l_j] + h.c.
\end{equation}
The reduced Higgs potential involving only the doublet higgs scalar is
\begin{equation}
V = m^2 \Phi^\dagger \Phi + {1 \over 2} \left( \lambda_1 - {{2\mu^2} \over 
M^2} \right) (\Phi^\dagger \Phi)^2,
\end{equation}
where $m$ is the mass of the higgs doublet. 
The last term comes from the exchange of $\xi$.  As $\phi^0$ acquires 
a nonzero vacuum expectation value $v$, we obtain Eq.~(\ref{min}) 
as we should, and 
the neutrino mass matrix becomes $$-2 f_{ij} \mu v^2 / M^2 = 2 f_{ij} u$$.

In models with right handed neutrinos lepton number is violated when
the right handed Majorana neutrinos decay as we discussed earlier 
[equation (\ref{N})]. The out-of-equilibrium condition is satisfied
if the masses of the lightest of the right handed neutrinos satisfy
$M_1 > 10^7$ GeV, which is obtained by solving the Boltzmann equations.
There are two sources of CP violation in this scenario :

\begin{figure}[hb]
\vskip 2in\relax\noindent\hskip -.3in\relax{\includegraphics{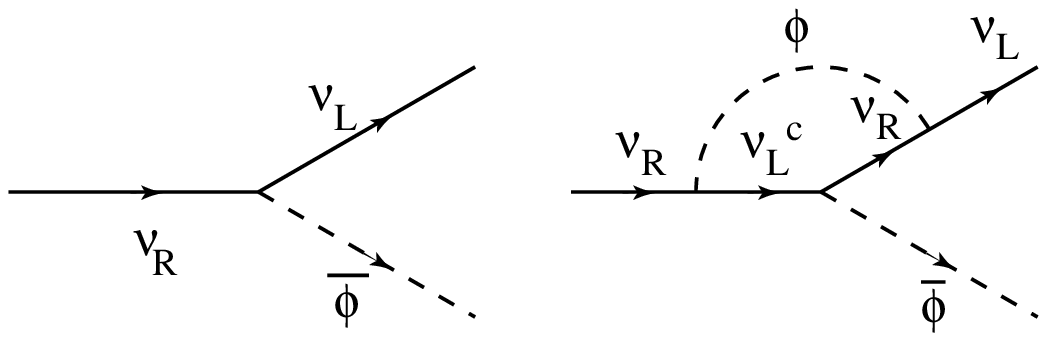}}
\caption{Tree and one loop vertex correction 
diagrams contributing to the generation of lepton asymmetry
in models with right handed neutrinos}
\end{figure}

\begin{itemize}
\item[$(i)$] vertex type diagrams which interferes with the tree level
diagram given by figure 2. 

\item[$(ii)$] self energy diagrams could interfere with the tree level
diagrams to produce CP violation analogous to CP violation in the 
$K^\circ \overline{K^\circ}$ oscillation as shown in figure 3. 
This type of CP violation has several interesting features, which
will be discussed in another talk at this meeting \cite{covi}.

\end{itemize}

In the case of self energy type $CP$ violation 
the amount of lepton asymmetry 
becomes large for a small mass difference between the two right-handed
heavy neutrinos ($M_1$ and $M_2$), and is given by 
\cite{lg3}:
\begin{equation}
\delta  =   {1 \over 8 \pi} {\cal C} \frac{M_1 M_2}{M_2^2 - M_1^2} 
\end{equation}
where
\begin{equation}
{\cal C} = - 
  {\rm Im} \left[ \sum_\alpha (h_{\alpha 1}^\ast h_{\alpha 2})
 \sum_\beta (h_{\beta 1}^\ast h_{\beta 2}) \right] \left( 
         \frac{1}{\sum_{\alpha} |h_{\alpha 1}|^2}
         + \frac{1}{\sum_{\alpha} |h_{\alpha 2}|^2 } \right)
\end{equation}
This   contribution   becomes   significant  when  the  two  mass
eigenvalues  are close to each other.  It  indicates  a resonance
like behaviour of the asymmetry if the two mass  eigenvalues  are
nearly  degenerate.  For very large values of the mass difference
this contribution becomes similar to the vertex correction. 
These two contributions add up to produce the final lepton asymmetry 
of the universe.

\begin{figure}[t]
\vskip 2.25in\relax\noindent\hskip -.3in\relax{\includegraphics{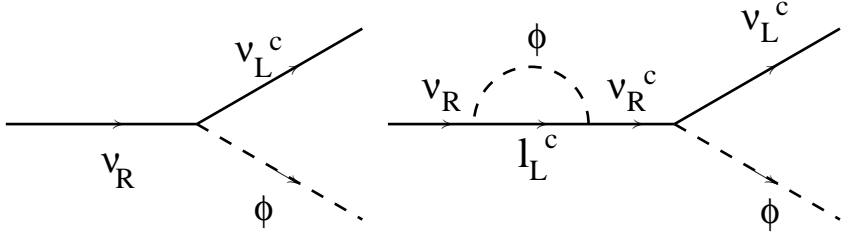}}
\caption{Tree and one loop self energy 
diagrams contributing to the generation of lepton asymmetry in models
with right handed neutrinos}
\end{figure}

In the triplet higgs scenario lepton number violation comes from the
decays of the triplet higgs $\xi_a$. Consider the decays of $\xi_a^{++}$, 
\begin{equation}
\xi_a^{++} \rightarrow \left\{ \begin{array} {l@{\quad}l} l_i^+ l_j^+ & 
(L = -2) \\ \phi^+ \phi^+ & (L = 0) \end{array} \right.
\end{equation}
The coexistence of the above two types of final states indicates the 
nonconservation of lepton number.  On the other hand, any lepton asymmetry 
generated by $\xi_a^{++}$ would be neutralized by the decays of $\xi_a^{--}$, 
unless CP conservation is also violated and the decays are out of thermal 
equilibrium in the early universe. In this case there are no
vertex corrections which can introduce CP violation. The only source
of CP violation is the self energy diagrams of figure 4. 

\begin{figure}[t]
\vskip 2.5in\relax\noindent\hskip -.5in\relax{\includegraphics{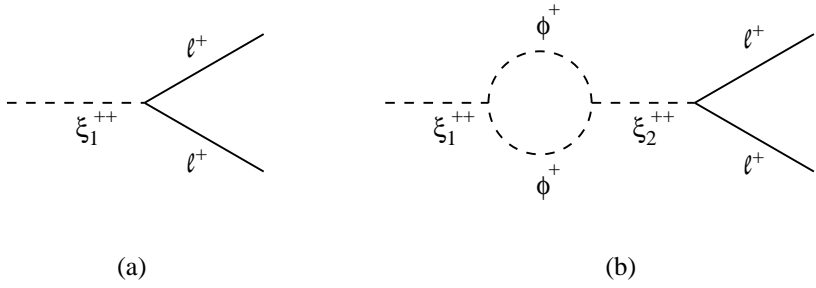}}
\caption{The decay of $\xi_1^{++} \to l^+ l^+$ at tree level (a) and in 
one-loop order (b).  A lepton asymmetry is generated by their interference
in the triplet higgs model for neutrino masses.}
\end{figure}

If there is only one $\xi$, then the 
relative phase between any $f_{ij}$ and $\mu$ can be chosen real.  Hence 
a lepton asymmetry cannot be generated.  With two $\xi$'s, even if there is 
only one lepton family, one relative phase must remain.  
As for the possible relative phases 
among the $f_{ij}$'s, they cannot generate a lepton asymmetry because they 
all refer to final states of the same lepton number.

In the presence of the one loop diagram, the mass matrix ${M_a}^2$ and
${M_a^*}^2$ becomes different. This implies that the rate of 
$\xi_b \to \xi_a$ no longer remains to be same as $\xi_b^* \to \xi_a^*$.
Since by $CPT$ theorem $\xi_b^* \to \xi_a^* \equiv \xi_a \to \xi_b$, 
what it means is that now $$ \Gamma[\xi_a \to \xi_b] \neq
\Gamma[\xi_b \to \xi_a] .$$
This is a different kind of CP violation compared to the CP violation
in models with right handed neutrinos. If we consider that the $\xi_2$
is heavier than $\xi_1$, then after $\xi_2$ decayed out the decay of
$\xi_1$ will generate an lepton asymmetry given by,
\begin{equation}
\delta \simeq 
{{Im \left[ \mu_1 \mu_2^* \sum_{k,l} f_{1kl} f_{2kl}^* \right]} \over 
{8 \pi^2 (M_1^2 - M_2^2)}} \left[ {{ M_1} \over 
\Gamma_1}  \right].
\end{equation}

In any of the above two scenarios with right handed neutrinos or with
triplet higgs, the lepton number evolves with time following the
Boltzmann equation in similar way. 
The lepton asymmetry $n_L$ thus generated through the CP violation
$\eta$ (where $\eta$ is the amount of CP violation in the model under
consideration), would evolve with time following the Boltzmann equation,
\begin{equation}
\frac{{\rm d}n_l}{{\rm d}t} + 3 H n_l = \eta
\Gamma_h [n_h - n_h^{eq} ] - \left(
\frac{n_l}{n_\gamma} \right) n_h^{eq} \Gamma_h
-2 n_\gamma n_l \langle \sigma |v| \rangle
\end{equation}
where $\Gamma_h$ is the thermally averaged decay rate of the heavy particle 
(whose number density is $n_h$).
The second term on the left side comes from the expansion of the universe,  
where $H = 1.66 g_*^{1/2} (T^2 / M_{Pl})$ is the Hubble constant.  
$n_\gamma$ is the photon  density and the
term  $\langle  \sigma  |v|  \rangle$   describes  the  thermally
averaged lepton number violating scattering cross section.  
The density of the heavy particle satisfies the Boltzmann equation,
\begin{equation}
\frac{{\rm d}n_h}{{\rm d}t} + 3 H n_h = - \Gamma_h (n_h - n_h^{eq})
\end{equation}

It is now convenient to use the dimensionless variable $x  =
{M_h}/{T}$ as well as the particle density per entropy density $Y_i
= {n_i}/{s}$, and the relation $t = {x^2}/{ 2H (x = 1) }$.  We also define 
the parameter $K \equiv \Gamma_h (x = 1) / H(x = 1)$ 
as a measure of the deviation from equilibrium. For $K << 1$ at 
$T \sim M_h$, the system is far from equilibrium; hence the last two terms 
responsible for the depletion of $n_l$ would be negligible.  With these 
simplifications and the above redefinitions, the Boltzmann equations 
effectively read:
\begin{equation}
\frac{{\rm d}Y_l}{{\rm d}x} = (Y_h - Y_h^{eq}) \eta K x, ~~~ 
\frac{{\rm d}Y_h}{{\rm d}x} = - (Y_h - Y_h^{eq}) K x.  
\end{equation}
In this limit $K << 1$, it is not difficult to obtain an 
asymptotic solution for $n_l$. Although the decay rate of 
$\psi_h$ is not fast enough to bring the number density $n_h$ to 
its equilibrium density, it is a good approximation to assume 
that the universe never goes far away from equilibrium. In 
other words, we can assume  
${\rm d}(Y_h - Y_h^{eq})/{\rm d}x = 0$ to 
get an asymptotic value for $Y_l$, given by $Y_l = n_l/s = \eta / g_*$. 
However, if $K > 1$, the terms which deplete $n_l$ 
dominate for some time and the lepton number density  
reaches its new asymptotic value, which is lower than the value 
it reaches in the out-of-equilibrium case. 
In this case although it is difficult to get 
an analytic solution of the Boltzmann equations, it is possible 
to get an approximate suppression factor, which is proportional to $K$.

In our earlier discussions we have considered the out-of-equilibrium
condition to be $$K < 1 ,$$ which gave us all the bounds on the 
Majorana masses of the left-handed neutrinos.
But as we can see from the above 
discussions, in models of leptogenesis, where a lepton number violation
is associated with a CP violation, the lepton number is not washed out
too fast. While in other models fast lepton number wash out any
preexisting asymmetry exponentially, in this case the depletion is
only linearly. As a result, in models of leptogenesis fast
lepton number violation may not wash out the primordial lepton 
asymmetry of the universe completely and the baryon asymmetry of the
universe may still be present after the lepton number violating
interaction goes out of equilibrium. 

\section{Summary }

Any fast lepton number violation in the universe can, in principle,
wash out baryon asymmetry of the universe. The survival of the
baryon asymmetry of the universe thus gives constraints on the left-handed
Majorana neutrino masses. There are constraints which are dependent
on models of the neutrino masses, but it is also possible to give
constraints from the lepton number violating interactions of $W^\pm$ 
due to the Majorana masses of the left-handed neutrinos directly. 
Although these constraints are independent of any models, these 
constraints also gets weakened in models of leptogenesis. 

\section*{Acknowledgement}

I would like to thank the Abdus Salam International Center for
Theoretical Physics, Trieste, Italy for providing financial 
assistance to attend this meeting, during my visit to the Center
as an Associate.

\end{document}